\documentclass[a4paper]{jpconf}
\usepackage{graphicx}
\usepackage{amssymb,amsmath,amsbsy}
\begin{document}
\title{Implications of symmetries in the scalar sector}

\author{Howard E Haber$^1$, O M Ogreid$^2$, P Osland$^3$ and M N Rebelo$^4$}

\address{$^1$ Santa Cruz Institute for Particle Physics, Santa Cruz, 
CA 95064, USA}
\address{$^2$ Western Norway University of Applied Sciences, 
Postboks 7030, N-5020 Bergen, Norway}
\address{$^3$ Department of Physics and Technology, University of Bergen,
Postboks 7803, N-5020 Bergen, Norway}
\address{$^4$ Departamento de F\'{\i}sica and Centro de F\'{\i}sica Te\'{o}rica de Part\'{\i}culas (CFTP), Instituto
Superior T\'{e}cnico, Av. Rovisco Pais, P-1049-001 Lisboa, Portugal}

\ead{haber@scipp.ucsc.edu, omo@hvl.no, Per.Osland@uib.no, rebelo@tecnico.ulisboa.pt (speaker) }

\begin{abstract}
Symmetries play a very important r\^ ole in Particle Physics. In extended scalar sectors, 
the existence of symmetries may permit the models to comply with the experimental constraints 
in a natural way, and at the same time reduce the number of free parameters.  There is a 
strong interplay among internal symmetries 
of the scalar potential, its CP properties and mass degeneracies of 
the physical scalars. Some of these aspects were discussed in this talk.
\end{abstract}

\section{Introduction}
There are strong motivations to extend the scalar sector of the Standard Model of 
Particle Physics by introducing more than one Higgs doublet. Models with two 
Higgs doublets have a very rich phenomenology and can give rise to spontaneous 
CP violation \cite{Lee:1973iz}, thus allowing to put electroweak symmetry breaking and CP 
violation on an equal footing. Two Higgs doublet models can also provide good dark matter 
candidates and can lead to new phenomena through Higgs mediated flavour changing neutral 
currents (HFCNC). Such models have been extensively studied in the literature. For 
reviews and references see for example \cite{Gunion:1989we,Branco:2011iw}. 

There are stringent experimental limits on HFCNC. One way of complying with
these limits is by imposing natural flavour conservation (NFC) through a $\mathbb{Z}_2$ symmetry
in such a way that all right handed isosinglet fermions of a given charge couple to no
more than one Higgs doublet \cite{Glashow:1976nt}.
In the case of NFC there are no tree level HCFNCs. Imposing a discrete symmetry in the
scalar sector of two Higgs doublet models eliminates
the possibility of having spontaneous CP violation. However, CP can still be spontaneously 
violated if one allows for the discrete symmetry to be softly broken \cite{Branco:1985aq}. 
An alternative way 
of suppressing HFCNC was proposed by Branco, Grimus and Lavoura \cite{Branco:1996bq}, and 
are the so-called BGL-type 
models \cite{Branco:1996bq,Botella:2009pq}, 
where it was shown that it is possible to have tree level FCNCs completely 
fixed by the $V_{\rm{CKM}}$ matrix as a result of an abelian symmetry. In these models, the 
suppression is given by the small entries of the $V_{\rm{CKM}}$ matrix with no new flavour 
structure. The same type of symmetry has also been extended to the leptonic sector 
\cite{Botella:2011ne}. 

Three Higgs doublets are also well motivated. In particular they allow for the 
presence of unbroken discrete symmetries and spontaneous CP violation. As the number 
of Higgs doublets grows so does the number of free parameters \cite{Olaussen:2010aq}. 
Symmetries  play a crucial r\^ ole in reducing the number of free parameters, thus increasing 
predictability. Finding whether or not there is spontaneous CP violation in multi-Higgs
models in the presence of symmetries may, in some cases, prove intricate. 
In ref.~\cite{Branco:1983tn} a simple relation is 
provided, allowing one to answer this question. However, in some cases it may not be 
straightforward to identify a matrix $U$ corresponding to a symmetry of the Lagrangian
that verifies the necessary relation. A simple alternative method based on the existence 
of scalar bases where only one doublet acquires a non-zero vacuum expectation value (vev) 
\cite{Donoghue:1978cj,Georgi:1978ri} was proposed in ref.~\cite{Ogreid:2017alh}.
This method may prove particularly useful as the complexity grows with the
number of doublets. 

This talk is based on several works by the present authors with special emphasis on 
refs.~\cite{Emmanuel-Costa:2016vej,Haber:2018iwr}. In section 2 we summarise the 
discussion on the interplay between symmetries and 
mass degeneracy in the case of two Higgs doublet models, as is presented 
in ref.~\cite{Haber:2018iwr}. 
In section 3 we discuss specific examples of three Higgs doublet models. In the latter case there is not yet 
a full study of all possible symmetries.

\section{Symmetries and mass degeneracy in two Higgs doublet models}
In the two Higgs doublet model (2HDM) the most general gauge-invariant renormalizable scalar 
potential is:
\begin{eqnarray}
\mathcal{V}&=& m_{11}^2 \Phi_1^\dagger \Phi_1+ m_{22}^2 \Phi_2^\dagger \Phi_2 -[m_{12}^2
\Phi_1^\dagger \Phi_2+{\rm h.c.}]
+\frac{1}{2} \lambda_1(\Phi_1^\dagger \Phi_1)^2+\frac{1}{2}  \lambda_2(\Phi_2^\dagger \Phi_2)^2
+\lambda_3(\Phi_1^\dagger \Phi_1)(\Phi_2^\dagger \Phi_2)
\nonumber\\
&&\quad
+\lambda_4( \Phi_1^\dagger \Phi_2)(\Phi_2^\dagger \Phi_1)
 +\left\{\frac{1}{2}  \lambda_5 (\Phi_1^\dagger \Phi_2)^2 +\big[\lambda_6 (\Phi_1^\dagger
\Phi_1) +\lambda_7 (\Phi_2^\dagger \Phi_2)\big] \Phi_1^\dagger \Phi_2+{\rm
h.c.}\right\}\,.\label{genpot}
\end{eqnarray}
The complete list of all possible symmetries of the 2HDM potential is known 
\cite{Deshpande:1977rw,Ivanov:2007de,Ferreira:2009wh,Ferreira:2010yh,Battye:2011jj,Pilaftsis:2011ed}. 
Table 1 gives the classification presented in ref.~\cite{Ferreira:2009wh}.
\begin{table}
\caption{Possible symmetries of the 2HDM scalar potential that are respected by the 
SU(2)$\times$U(1)$_{\rm Y}$ gauge kinetic term of the scalar fields.  The corresponding 
symmetry transformation laws are given in a basis where the symmetry is manifest. Table taken 
from ref.~\cite{Haber:2018iwr}}
\begin{center}
\begin{tabular}{|cccc|} 
\hline
\rule{0pt}{\normalbaselineskip} 
symmetry & \hspace{0.5in} transformation law  & \hspace{0.7in} \phantom{transformation law}  & \\[2pt]
\hline 
\rule{0pt}{\normalbaselineskip} 
$\mathbb{Z}_2$ & $\Phi_1\to \Phi_1$ &  $\Phi_2\to -\Phi_2$ &\\
U(1)  & $\Phi_1\to  \Phi_1$ &  $\Phi_2\to e^{2i\theta}\Phi_2$ &\\
SO(3) & $\Phi_a\to U_{ab}\Phi_b$ & $U\in {\rm U}(2)/{\rm U}(1)_{\rm Y}$& (\mbox{for $a$, $b=1,2$} )\\
GCP1 & $\Phi_1\to\Phi_1^\ast$ &  $\Phi_2\to\Phi_2^\ast$ & \\
GCP2 & $\Phi_1\to\Phi_2^\ast$ &  $\Phi_2\to -\Phi_1^\ast$ & \\
GCP3 & $\Phi_1 \rightarrow \Phi_1^\ast \cos\theta+\Phi_2^\ast \sin\theta$ &
$\Phi_2 \rightarrow -\Phi_1^\ast \sin\theta+\Phi_2^\ast\cos\theta$ & (for $0<\theta< \frac{1}{2}\pi$)  \\[2pt]
\hline\hline 
\rule{0pt}{\normalbaselineskip} 
$\Pi_2$& $\Phi_1\to\Phi_2$ & $\Phi_2\to \Phi_1$ & \\[2pt]
\hline 
\end{tabular}
\end{center}
\end{table}
If the scalar potential respects one of the symmetries of Table 1, the coefficients of the scalar 
potential are constrained according to Table 2. 
\begin{table}
\caption{\small Impact of the symmetries defined in Table 1 on the coefficients of the 
2HDM scalar potential in a 
basis where the symmetry is manifest.  A short dash indicates the absence of a constraint. Table taken 
from ref.~\cite{Haber:2018iwr}}
\begin{center}
\begin{tabular}{|ccccccccccccc|}
\hline
\rule{0pt}{\normalbaselineskip} 
symmetry & $m_{11}^2$ &  $m_{22}^2$ & $m_{12}^2$ & $\lambda_1$ &
 $\lambda_2$ & $\lambda_3$ & $\lambda_4$ &
$\mbox{Re}\lambda_5$ & $\mbox{Im}\lambda_5$ & $\lambda_6$ & $\lambda_7$ & \\[2pt]
\hline
\rule{0pt}{\normalbaselineskip} 
$\mathbb{Z}_2$ & - & -   & 0 & - 
   &  -  &  -  &  -  &  - &  -
   & 0 & \phantom{-} 0  & \\
U(1) & -  & -    & 0  & - 
 & -  & - &  -  & 0 &
0 & 0 & \phantom{-} 0 & \\
SO(3)  & - &   $ m_{11}^2$ & 0 & - 
   & $\lambda_1$ &- & $\lambda_1 - \lambda_3$ & 0 &
0 & 0 & \phantom{-} 0  & \\
GCP1   & -  & -  & real & - 
 & -  &  - &  -  &
- & 0 & real &\phantom{-} real  & \\
GCP2   & -  & $m_{11}^2$ & 0  & - 
  & $\lambda_1$ &  - &  -  & - &-
   &  -  & $-\lambda_6$ & \\
GCP3  & -   & $m_{11}^2$ & 0 & - 
   & $\lambda_1$ &  -  &  - &
$\lambda_1 - \lambda_{3}-\lambda_4$ & 0 & 0 & \phantom{-} 0 & \\[2pt]
\hline\hline
\rule{0pt}{\normalbaselineskip} 
$\Pi_2$  & -  & $ m_{11}^2$ & real & - &
    $ \lambda_1$ &  - &   -&  
- & 0 & - & $\lambda_6^*$ &
\\
$\mathbb{Z}_2\oplus\Pi_2$ & -  & $ m_{11}^2$ & $0$ & - &
    $ \lambda_1$ &  - &   - &  
- & 0 & 0 &\phantom{-} 0 &
\\
U(1)$\oplus\Pi_2$ & -  & $ m_{11}^2$ & $0$ & -  &  $\lambda_1$  & - & - & 0 & 0 & 0 & \phantom{-} 0 &
\\[2pt]
\hline
\end{tabular}
\end{center}
\end{table}
Two different complementary approaches were followed in ref.~\cite{Haber:2018iwr} in 
order to find out under what conditions there is exact degeneracy of scalar masses 
without an artificial fine-tuning of the parameters of the scalar potential. 
The first one relies on deriving  directly from the potential the conditions that yield 
mass degeneracies and checking afterwards whether these conditions correspond to a symmetry. 
The second one takes as  starting point the set of all possible symmetries together with 
the constraints that these symmetries impose on the coefficients of the potential in the basis 
where the symmetry is imposed. Both approaches obviously lead to the same conclusions.
In the 2HDM a natural mass degeneracy can only arise in the case of the inert 2HDM, in which
there is an exact discrete $\mathbb{Z}_2$ symmetry left unbroken by the vacuum. In this
case, the two neutral scalars $H$ and $A$ that reside in the inert doublet are mass
degenerate provided that $\lambda_5 = 0$.
Note that we are assuming that the field that is odd under $\mathbb{Z}_2$ 
acquires zero vacuum expectation 
value. This corresponds to the second row of Tables 1 and 2 with an unbroken U(1) symmetry.
Furthermore, the global U(1) symmetry responsible for the degeneracy of the masses of $H$ and $A$
is preserved by the interactions of the scalars with the vector bosons.
The SO(3) symmetric case (third row of Tables 1 and 2), again with the symmetry imposed in the basis
where one of the doublets has zero vev, is even more constrained than 
the U(1) symmetric one and, in this case, the degeneracy is preserved with both scalars being massless 
($m_H=m_A = 0$). The presence of two massless states is the consequence of the spontaneous breaking
of the SO(3) symmetry. In the next three rows three different cases of CP symmetries are listed,
GCP1, GCP2, GCP3. The necessary conditions in each of these cases have been expressed in terms
masses and physical couplings in ref.~\cite{Grzadkowski:2019nwa}, published in these Proceedings.

\section{The case of three Higgs doublet models. Particular examples}

As mentioned before, in the case of three Higgs doublets there is not yet a full classification of all 
possible symmetries of the scalar potential. In this section we discuss two particular cases. 
In the first case we analyse the possibility of having spontaneous CP violation in a three Higgs doublet 
model with real coefficients where an $S_3$ symmetry is imposed. This discussion is based on 
ref.~\cite{Emmanuel-Costa:2016vej}. In the second case we discuss some properties of
a three Higgs doublet model proposed by Ivanov and Silva \cite{Ivanov:2015mwl} with the striking property
of having no real scalar basis and still preserving CP.

\subsection{Spontaneous CP violation in the $S_3$ symmetric scalar sector}

The $S_3$ symmetric three Higgs doublet scalar potential has been studied by several authors both in
terms of its irreducible doublet-singlet representation, $h_1$, $h_2$ and $h_S$ representation \cite{Pakvasa:1977in,Kubo:2004ps,Teshima:2012cg,Das:2014fea}
as well as in terms of a reducible triplet,
$\phi_1$, $\phi_2$ and $\phi_3$ \cite{Derman:1978rx,Derman:1979nf}.
In ref.~\cite{Emmanuel-Costa:2016vej} a complete list of all possible vacua, starting from a scalar potential with real coefficients
was given together with the set of constraints on the parameters of the potential that must be obeyed in each case.

In terms of the reducible triplet fields the potential has the form \cite{Derman:1978rx}:
\begin{equation}
\label{Eq:pot-original}
V_2=-\lambda\sum_{i}\phi_i^\dagger\phi_i +\frac{1}{2}\gamma\sum_{i<j}[\phi_i^\dagger\phi_j+ \mbox{h.c.} ],\\
\end{equation}

\begin{eqnarray}
V_4&=A\sum_{i}(\phi_i^\dagger\phi_i)^2
+\sum_{i<j}\{C(\phi_i^\dagger\phi_i)(\phi_j^\dagger\phi_j)
+\overline C (\phi_i^\dagger\phi_j)(\phi_j^\dagger\phi_i) 
+\frac{1}{2} D[(\phi_i^\dagger\phi_j)^2+ \mbox{h.c.}]\} \nonumber \\
&+\frac{1}{2} E_1\sum_{i\neq j}[(\phi_i^\dagger\phi_i)(\phi_i^\dagger\phi_j)+  \mbox{h.c.} ]
+\sum_{i\neq j\neq k,j<k}
\{\frac{1}{2} E_2[(\phi_i^\dagger\phi_j)(\phi_k^\dagger\phi_i)+ \mbox{h.c.} ] 
\nonumber \\
&+\frac{1}{2} E_3[(\phi_i^\dagger\phi_i)(\phi_k^\dagger\phi_j)+ \mbox{h.c.}] 
+\frac{1}{2} E_4[(\phi_i^\dagger\phi_j)(\phi_i^\dagger\phi_k)+ \mbox{h.c.} ]\}.
\label{29ab}
\end{eqnarray}

Whereas in terms of the $S_3$ singlet and doublet fields it can be written as 
\cite{Kubo:2004ps,Teshima:2012cg,Das:2014fea}:
\begin{equation}
\label{Eq:V-DasDey}
V_2=\mu_0^2 h_S^\dagger h_S +\mu_1^2(h_1^\dagger h_1 + h_2^\dagger h_2),
\end{equation}

\begin{eqnarray}
V_4&=
\lambda_1(h_1^\dagger h_1 + h_2^\dagger h_2)^2 
+\lambda_2(h_1^\dagger h_2 - h_2^\dagger h_1)^2
+\lambda_3[(h_1^\dagger h_1 - h_2^\dagger h_2)^2+(h_1^\dagger h_2 + h_2^\dagger h_1)^2]
\nonumber \\
&+ \lambda_4[(h_S^\dagger h_1)(h_1^\dagger h_2+h_2^\dagger h_1)
+(h_S^\dagger h_2)(h_1^\dagger h_1-h_2^\dagger h_2)+  \mbox{h.c.}] 
+\lambda_5(h_S^\dagger h_S)(h_1^\dagger h_1 + h_2^\dagger h_2) \nonumber \\
&+\lambda_6[(h_S^\dagger h_1)(h_1^\dagger h_S)+(h_S^\dagger h_2)(h_2^\dagger h_S)] 
+\lambda_7[(h_S^\dagger h_1)(h_S^\dagger h_1) + (h_S^\dagger h_2)(h_S^\dagger h_2) + \mbox{h.c.}]
\nonumber \\
&+\lambda_8(h_S^\dagger h_S)^2.
\label{Eq:V-DasDey-quartic}
\end{eqnarray}

Table 3 lists all possible complex vacua. In ref.~\cite{Emmanuel-Costa:2016vej} the set of 
constraints required for each solution is given. In the discussion of whether or not there is spontaneous 
CP violation (SCPV) for each of the solutions, it was shown that the parameter $\lambda_4$ plays a very important 
r\^ ole.  In fact for $\lambda_4 = 0$ there is an additional SO(2) symmetry. The results are summarised in Table 4
indicating whether or not  $\lambda_4 $ is required to be zero (an X means that it can be different from zero). 
There is no spontaneous CP violation in any of the cases with $\lambda_4 = 0$. The simple method described 
in \cite{Ogreid:2017alh} proved very useful in this discussion. In this framework there are also mass degeneracies 
among scalar fields, however, the identification of such degeneracies for the different vacuum solutions was not done 
in our previous work.

\begin{table}
\caption{Complex vacua. Notation: $\epsilon=1$ and $-1$ for C-III-d and C-III-e, respectively;
$\xi=\sqrt{-3\sin 2\rho_1/\sin2\rho_2}$,
$\psi=\sqrt{[3+3\cos (\rho_2-2 \rho _1)]/(2\cos\rho_2)}$. With the necessary constraints the vacua labelled with an asterisk ($^\ast$) are in fact real.}
\label{Table:complex}
\begin{center}
\begin{tabular}{|c|c|c|}
\hline\hline
& IRF (Irreducible Rep.)& RRF  (Reducible Rep.) \\
\hline
& $w_1,w_2,w_S$ & $\rho_1,\rho_2,\rho_3$  \\
\hline
\hline
C-I-a & $\hat w_1,\pm i\hat w_1,0$ & 
$x, xe^{\pm\frac{2\pi i}{3}}, xe^{\mp\frac{2\pi i}{3}}$ \\
\hline
\hline
C-III-a & $0,\hat w_2e^{i\sigma_2},\hat w_S$ & $y, y, xe^{i\tau}$  \\
\hline
C-III-b & $\pm i\hat w_1,0,\hat w_S$ & $x+iy,x-iy,x$  \\
\hline
C-III-c & $\hat w_1 e^{i\sigma_1},\hat w_2e^{i\sigma_2},0$ 
& $xe^{i\rho}-\frac{y}{2}, -xe^{i\rho}-\frac{y}{2}, y$  \\
\hline
C-III-d,e & $\pm i \hat w_1,\epsilon\hat w_2,\hat{w}_S$ & $xe^{ i\tau},xe^{- i\tau},y$ \\
\hline
C-III-f & $\pm i\hat w_1 ,i\hat w_2,\hat{w}_S$ 
& $re^{i\rho}\pm ix,re^{i\rho}\mp ix,\frac{3}{2}re^{-i\rho}-\frac{1}{2}re^{i\rho}$ \\
\hline
C-III-g & $\pm i\hat w_1,-i\hat w_2,\hat{w}_S$ 
& $re^{-i\rho}\pm ix,re^{-i\rho}\mp ix,\frac{3}{2}re^{i\rho}-\frac{1}{2}re^{-i\rho}$ \\
\hline
C-III-h & $\sqrt{3}\hat w_2 e^{i\sigma_2},\pm\hat w_2 e^{i\sigma_2},\hat{w}_S$ 
& $xe^{i\tau} , y , y$ \\
& & $y, xe^{i\tau},y$ \\
\hline
C-III-i & $\sqrt{\frac{3(1+\tan^2\sigma_1)}{1+9\tan^2\sigma_1}}\hat w_2e^{i\sigma_1},$ 
& $x, ye^{i\tau},ye^{-i\tau}$ \\
& $\pm\hat w_2e^{-i\arctan(3\tan\sigma_1)},\hat w_S$ 
& $ye^{i\tau}, x, ye^{-i\tau}$ \\
\hline
\hline
C-IV-a$^\ast$ & $\hat w_1e^{i\sigma_1},0,\hat w_S$ & $re^{i\rho}+x, -re^{i\rho}+x,x$ \\
\hline
C-IV-b & $\hat w_1,\pm i\hat w_2,\hat w_S$ 
& $re^{i\rho}+x, -re^{-i\rho}+x, -re^{i\rho}+re^{-i\rho}+x$ \\
\hline
C-IV-c & $\sqrt{1+2\cos^2\sigma_2}\hat w_2,$ &  $re^{i\rho}+r\sqrt{3(1+2\cos^2\rho)}+x$, \\
& $\hat w_2e^{i\sigma_2},\hat w_S$ 
& $re^{i\rho}-r\sqrt{3(1+2\cos^2\rho)}+x,-2re^{i\rho}+x$ \\
\hline
C-IV-d$^\ast$ & $\hat w_1e^{i\sigma_1},\pm\hat w_2e^{i\sigma_1},\hat w_S$ & $r_1e^{i\rho}+x, (r_2-r_1)e^{i\rho}+x,-r_2e^{i\rho}+x$ \\
\hline
C-IV-e & $\sqrt{-\frac{\sin 2\sigma_2}{\sin 2\sigma_1}}\hat w_2e^{i\sigma_1},$ & $re^{i\rho_2}+re^{i\rho_1}\xi+x, re^{i\rho_2}-re^{i\rho_1}\xi+x,$  \\
& $\hat w_2e^{i\sigma_2},\hat w_S$ & $-2re^{i\rho_2}+x$ \\
\hline
C-IV-f & $\sqrt{2+\frac{\cos \left(\sigma _1-2 \sigma _2\right)}{\cos\sigma_1}}\hat w_2e^{i\sigma_1},$ & $re^{i\rho_1}+re^{i\rho_2}\psi+x$, \\
& $\hat w_2e^{i\sigma_2},\hat w_S$ &$re^{i\rho_1}-re^{i\rho_2}\psi+x, -2re^{i\rho_1}+x$ \\
\hline
\hline
C-V$^\ast$ & $\hat w_1e^{i\sigma_1},\hat w_2e^{i\sigma_2},\hat w_S$ & $xe^{i\tau_1},ye^{i\tau_2},z$ \\
\hline
\end{tabular}
\end{center}
\end{table}

\begin{table}
\caption{Spontaneous CP violation (SCPV)}
\label{Table:CPV}
\begin{center}
\begin{tabular}{|c|c|c||c|c|c||c|c|c|}
\hline
Vacuum  &  $\lambda_4$ & SCPV & Vacuum  &  $\lambda_4$ &  SCPV 
& Vacuum &  $\lambda_4$ &  SCPV\\
\hline
C-I-a & X & no & C-III-f,g & 0 & no & C-IV-c & X & yes   \\
C-III-a & X & yes  & C-III-h & X & yes  & C-IV-d & 0 & no \\
C-III-b & 0 & no  & C-III-i & X & no & C-IV-e & 0 & no \\
C-III-c & 0 & no  & C-IV-a & 0 & no & C-IV-f & X & yes \\
C-III-d,e & X & no  & C-IV-b & 0 & no & C-V & 0 & no\\
\hline
\end{tabular}
\end{center}
\end{table}

\subsection{A CP-conserving multi-Higgs model with irremovable complex coefficients} 
Ivanov and Silva (IS) proposed a three Higgs doublet scalar potential \cite{Ivanov:2015mwl}
with irremovable complex coefficients that conserves CP. The IS potential has the form:
\begin{equation}
V = V_0 + V_1
\end{equation}
where:
\begin{eqnarray}
V_0&=-m_{11}^2(\phi_1^\dagger\phi_1)-m_{22}^2(\phi_2^\dagger\phi_2+\phi_3^\dagger\phi_3)
+\lambda_1(\phi_1^\dagger\phi_1)^2  
+\lambda_2[(\phi_2^\dagger\phi_2)^2+(\phi_3^\dagger\phi_3)^2]+\lambda_3^\prime(\phi_2^\dagger\phi_2)(\phi_3^\dagger\phi_3)\nonumber \\
& \qquad
+\lambda_3(\phi_1^\dagger\phi_1)[(\phi_2^\dagger\phi_2)+(\phi_3^\dagger\phi_3)] 
+\lambda_4^\prime(\phi_2^\dagger\phi_3)(\phi_3^\dagger\phi_2)
+\lambda_4[(\phi_1^\dagger\phi_2)(\phi_2^\dagger\phi_1)+(\phi_1^\dagger\phi_3)(\phi_3^\dagger\phi_1)], 
\end{eqnarray}
with all parameters real, due to hermiticity and
\begin{eqnarray}
V_1&=\lambda_5(\phi_3^\dagger\phi_1)(\phi_2^\dagger\phi_1)
+\frac{1}{2}\lambda_6[(\phi_2^\dagger\phi_1)^2-(\phi_1^\dagger\phi_3)^2]
+\lambda_8(\phi_2^\dagger\phi_3)^2 \nonumber \\
& +\lambda_9(\phi_2^\dagger\phi_3)[(\phi_2^\dagger\phi_2)-(\phi_3^\dagger\phi_3)] +{\rm h.c.}
\end{eqnarray}
with $\lambda_5$, $\lambda_6$ real and $\lambda_8$, $\lambda_9$ complex. This potential is 
fixed by the following CP symmetry:
\begin{equation} \label{Eq:X-def}
\phi_i \to W_{ij}\phi_j^\ast, \quad
W = \left( 
\begin{array}{ccc}
1 &  0 &  0 \\
0 &  0 &  i \\
0 & -i &  0
\end{array}
\right)
\end{equation}
Notice that this symmetry requires $\lambda_5$ to be real. On the other hand $\lambda_6$
can be made real by an appropriate rephasing of the fields $\phi_2$ and $\phi_3$. 
There exists a range of parameters of the scalar potential corresponding to a vacuum that 
conserves the above symmetry, i.e., $(v_1,v_2,v_3)=(v,0,0)$.
As pointed out by Ivanov and Silva \cite{Ivanov:2015mwl} the transformation given above 
is of order 4 since only after applying it four times are we back to the identity 
transformation and therefore was called a CP4 transformation. Notice that $WW^\ast = \mbox{diag} (1,-1,-1)$ is
different from the identity. This may seem paradoxical since one expects that two
consecutive CP transformations would correspond to the identity. However, it should be pointed out 
that this happens because this CP transformation combines the usual CP transformation 
for a single scalar doublet with an internal symmetry of the potential.
From the point of view of spacetime we can only interpret this transformation as a CP transformation
if we assume that each time it is applied $\vec{x}$ goes into -$\vec{x}$, so that
after applying it twice we are simply left with an internal symmetry transformation of the
scalar potential.

It was pointed out in \cite{Haber:2018iwr} that there is a simpler form for the IS potential
with no $\lambda_5$ term, with the term in $\lambda_6$ changed into 
$\frac{1}{2}\lambda_6^\prime [(\phi_2^{\prime \dagger}\phi_1)^2+(\phi_1^\dagger\phi_3^\prime)^2]$,
with $\lambda_6^\prime$ still real, and all other terms keeping the same form, though
with different values for the coefficients. Furthermore, it is still possible to
make either  $\lambda_8^\prime$ or $\lambda_9^\prime$ real. This means that
the $\lambda_5$ term is made redundant and only one coefficient remains complex. This is achieved 
by a series of unitary transformations under which $\phi_1$ does not mix with the other
two doublets. As a result the new doublets with indices two and three still have zero vevs. Furthermore,
in this basis all fields appearing in $\phi_2^\prime$ and $\phi_3^\prime$ are already mass eigenstates
and do not mix with the  Standard Model-like Higgs boson. In addition, the symmetry of the IS model
potential is also responsible for the existence of pairwise mass degeneracy among all the
fields appearing in $\phi_2^\prime$ and $\phi_3^\prime$. Once again, there is a strong interplay 
between symmetry and mass degeneracies.

The IS model has the curious property that there is no CP transformation of order two (CP2) under 
which the potential is invariant. It also has the curious property that there is no real basis even 
though CP is conserved. In ref.~\cite{Haber:2018iwr}, we have checked up to three-loop order 
in the IS model that there is a cancellation of the CP violating form factors of 
effective ZZZ and ZWW vertices. We have also identified a physical quartic scalar interaction
that is consistent with the CP4 symmetry but would vanish for a potential
of the same form as the IS potential written in terms of real coefficients.

\section{Conclusions}
Symmetries play a crucial r\^ ole in Particle Physics. The imposition of symmetries on the scalar
potential, with multi-Higgs doublets, leads to degeneracies in the masses of the physical 
scalars and has implications on its CP properties. A full study of such implications
in the case of models with three Higgs doublets has not yet been performed.

\section*{Acknowledgments}
M. N. R. thanks the organisers of DISCRETE 2018 and of the 14th Vienna Central European Seminar (VCES) for 
the very stimulating Meetings and the hospitality. 
H.E.H.~is supported in part by the U.S. Department of Energy grant
number DE-SC0010107, and in part by the grant H2020-MSCA-RISE-2014
No.~645722 (NonMinimalHiggs).
P.O.~is supported by the Research Council of Norway.
The work of M.N.R. was partially supported by Funda\c c\~ ao 
para  a  Ci\^ encia e a Tecnologia  (FCT, Portugal)  through  the  projects  
CFTP-FCT Unit  777(UID/FIS/00777/2013), (UID/FIS/00777/2019), CERN/FIS-PAR/0004/2017 and 
PTDC/FIS-PAR/29436/2017 which are  partially  funded  through  POCTI  (FEDER),  COMPETE,  QREN  and  EU.
M.N.R. benefited from discussions that took place at the University of Warsaw during visits supported  by  
the  HARMONIA  project  of  the  National  Science  Centre,  Poland,  under  
contract UMO-2015/18/M/ST2/00518 (2016–2019).  M.N.R. also thanks the University of Bergen where 
collaboration visits took place.

\end{document}